\documentclass[fleqn,twoside]{article}

\usepackage{amsmath}
\usepackage{espcrc2}
\usepackage{graphicx}
\usepackage[figuresright]{rotating}

\def\slashchar#1{\setbox0=\hbox{$#1$}
   \dimen0=\wd0 \setbox1=\hbox{/} \dimen1=\wd1
   \ifdim\dimen0>\dimen1 \rlap{\hbox to \dimen0{\hfil/\hfil}} #1
   \else  \rlap{\hbox to \dimen1{\hfil$#1$\hfil}} / \fi}
\def\tstrut{\vrule height2.5ex depth0pt width0pt} 

\title{Nuclear Many-Body Theory of Electroweak Interactions with Nuclei at
       Intermediate Energies}

\author{J. Nieves\address[DFM]{Departamento de F\'\i sica Moderna, \\ 
        Universidad de Granada, E-18071 Granada, Spain},
        J. E. Amaro\addressmark,
        M. Valverde\addressmark}
       
\begin{document}

\begin{abstract}
The Quasi-Elastic (QE) contribution of the nuclear inclusive electron
model developed in reference~\cite{GNO97} is extended to the study of
electroweak Charged Current (CC) induced nuclear reactions at
intermediate energies of interest for future neutrino oscillation
experiments. Long range nuclear (RPA) correlations, Final State
Interaction (FSI) and Coulomb corrections are included within the
model. RPA correlations are shown to play a crucial role in the whole
range of neutrino energies, up to 500 MeV, studied in this work.
Predictions for inclusive muon capture for different nuclei through
the Periodic Table and for the reactions $^{12}$C $(\nu_\mu,\mu^-)X$ and
$^{12}$C $(\nu_e,e^-)X$ near threshold are also given.
\vspace{1pc}
\end{abstract}

\maketitle

\section{INTRODUCTION}
Any model aiming at describing the interaction of neutrinos with
nuclei should be firstly tested against the existing data on the
interaction of real and virtual photons with nuclei. At intermediate
energies, nuclear excitation energies ranging from about 100 MeV to
500 or 600 MeV, three different contributions should be taken into
account: i) QE processes, ii) pion production and two body processes
from the QE region to that beyond the $\Delta(1232)$ resonance peak,
and iii) double pion production and higher nucleon resonance degrees
of freedom induced processes.  The model developed in
Refs.~\cite{GNO97} (inclusive electro--nuclear reactions)
and~\cite{CO92} (inclusive photo--nuclear reactions) has been
successfully compared with data at intermediate energies and it
systematically includes the three type of contributions mentioned
above.  Nuclear effects are computed starting from a Local Fermi Gas
(LFG) picture of the nucleus, an accurate approximation to deal with
inclusive processes which explore the whole nuclear
volume~\cite{CO92}, and their main features, expansion parameter and
all sort of constants are completely fixed from previous
hadron-nucleus studies (pionic atoms, elastic and inelastic
pion-nucleus reactions, $\Lambda-$ hypernuclei, etc...)~\cite{pion}.
Thus, and besides the photon coupling constants determined in the
vacuum, the model of Refs.~\cite{GNO97} and \cite{CO92} has no free
parameters, and thus the results presented in these two references are
predictions deduced from the nuclear microscopic framework developed
in Refs.~\cite{OTW82} and \cite{pion}. In this talk, we extent the
nuclear inclusive QE electron scattering model of Ref.~\cite{GNO97},
including the axial CC degrees of freedom, to describe neutrino and
antineutrino induced nuclear reactions in the QE region.  Detailed
formulae and results can be found in \cite{NAV04}.

\begin{table*}[t]
 \caption{\footnotesize Experimental and theoretical flux averaged
 $^{12}{\rm C}(\nu_\mu,\mu^-)X$ cross sections. The notation for the
  theoretical predictions is the same as in Fig.~\protect\ref{fig:lsnd}.}
 \label{tab:lsnd}
 \begin{tabular}{lcccccc}\hline\tstrut 
  &\multicolumn{3}{c}{Theory} & \multicolumn{3}{c}{ 
Exp (LSND)~\protect~\cite{Al95} }
 \\\hline\tstrut
 & LDT & Pauli & RPA &
 1995&1997  &
 2002 \\\hline\tstrut
 $\overline{\sigma}$ [10$^{-40}$ cm$^2$] & ~66.1~ & ~20.7~ & ~11.9~ & ~$8.3
 \pm 0.7 \pm 1.6$~ & ~$11.2 \pm 0.3 \pm 1.8$ ~& ~$10.6 \pm 0.3 \pm 1.8$~ \\
 \hline 
\end{tabular}
\end{table*}
\begin{table*}[t]
 \caption{\footnotesize Experimental and theoretical flux averaged
 $^{12}{\rm C}(\nu_e,e^-)X$ cross sections. The notation for the
  theoretical predictions is the same as in Fig.~\protect\ref{fig:lsnd}.}
 \label{tab:eincl}
\begin{tabular}{lcccccc}\hline\tstrut 
  &\multicolumn{3}{c}{Theory} & \multicolumn{3}{c}{ Exp~\protect~\cite{KARMEN}  }
 \\\hline\tstrut
 & LDT & Pauli & RPA &
 KARMEN&LSND  &
 LAMPF \\\hline\tstrut
 $\overline{\sigma}$ [10$^{-40}$ cm$^2$] & ~5.97~ & ~0.19~ & ~0.14~ & ~$0.15
 \pm 0.01 \pm 0.01$~ & ~$0.15 \pm 0.01 \pm 0.01$ ~& ~$0.141 \pm 0.023$~ \\
 \hline \end{tabular}
\end{table*}

\section{INCLUSIVE CROSS SECTION }

We will expose here the general formalism focusing on the neutrino CC
reaction. The generalization to antineutrino CC reactions or muon
capture is straightforward.
\begin{figure}[htb]
\centerline{\includegraphics[height=4cm]{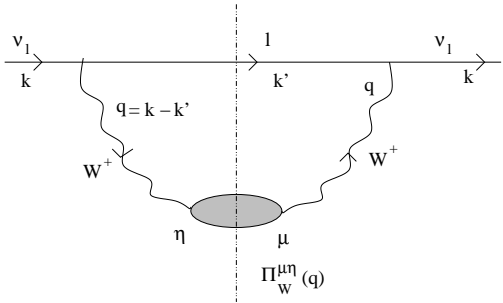}}
\vspace{-1cm}
\caption{\footnotesize Diagrammatic representation of the neutrino
  selfenergy in nuclear matter. }\label{fig:fig1}
\end{figure}
In the laboratory frame, the differential cross section for the
process $\nu_l (k) +\, A_Z \to l^- (k^\prime) + X $ reads:
\begin{equation}
\frac{d^2\sigma}{d\Omega(\hat{k^\prime})dE^\prime_l} =
\frac{|\vec{k}^\prime|}{|\vec{k}~|}\frac{G^2}{4\pi^2} 
L_{\mu\sigma}W^{\mu\sigma} 
\end{equation}
with $L$ and $W$ the leptonic and hadronic tensors, respectively. On
the other hand, the inclusive CC nuclear cross section is related to
the imaginary part of the neutrino self-energy (see
Fig.~\ref{fig:fig1}) in the medium by:
\begin{equation}
\sigma = - \frac{1}{|\vec{k}|} \int {\rm Im} \Sigma_\nu (k;\rho(r)) d^3r
\label{eq:sSrel}
\end{equation}
We get ${\rm Im} \Sigma_\nu$
by following the prescription of the Cutkosky's rules: in this case we
cut with a vertical straight line (see Fig.~\ref{fig:fig1}) the
intermediate lepton state and those implied by the $W$-boson
polarization.  Those states are placed on shell by taking the
imaginary part of the propagator , self-energy, etc. We obtain for
$k^0 > 0$
\begin{equation}
  {\rm Im} \Sigma_\nu(k) = \frac{8G\,\Theta(q^0)}{\sqrt 2 M^2_W}\int \frac{d^3
  k^\prime}{(2\pi)^3 }\frac{ {\rm Im}\left\{ \Pi^{\mu\eta}_W 
L_{\eta\mu} \right\}}{2E^{\prime}_l}  
\label{eq:ims}
\end{equation}
and thus, the hadronic tensor  is basically
an integral over the nuclear volume of the $W-$selfenergy
($\Pi_W^{\mu\nu}(q;\rho)$)  inside  the nuclear medium. We can
then take into account the  in-medium effects (such as $W-$absorption by pairs
of nucleons,  pion
production, delta resonances,...)  by including the correspondent
diagram in the shaded loop of Fig.~\ref{fig:fig1}. Further details can
be found in Ref.~\cite{GNO97}.

\section{QE CONTRIBUTION TO $\Pi_W^{\mu\nu}(q;\rho)$}
The virtual $W^+$ can be absorbed by one nucleon (neutron) leading to the QE
peak of the nuclear response function. Such a contribution corresponds to a
1p1h nuclear excitation.
We consider a structure of the $V-A$ type for the $W^+pn$ vertex, and use
PCAC and
invariance under G-parity  to relate the
pseudoscalar form factor to the axial one and to discard a term of the
form $(p^\mu+p^{\prime \mu})\gamma_5$ in the axial sector,
respectively.  Invariance under time reversal guarantees that all form
factors are real. Besides, and thanks to isospin symmetry, the vector
form factors are related to the electromagnetic ones. 
We find 
\begin{equation}
\begin{split}
W^{\mu\nu}&(q) 
=\frac{\cos^2\theta_C}{2M^2} \int_0^\infty dr r^2 \Big \{
\Theta(q^0) \int \frac{d^3p}{4\pi^2} \frac{M}{E(\vec{p})}
 \\
& \frac{M}{E(\vec{p}+\vec{q})}  
 \Theta(k_F^n(r)-|\vec{p}~|) \Theta(|\vec{p}+\vec{q}~|-k_F^p(r))
 \\
 &\delta(q^0 +
E(\vec{p}) -E(\vec{p}+\vec{q}~))   A^{\nu\mu}(p,q)|_{p^0=E(\vec{p})~} 
\Big \}
\label{eq:thadron}
\end{split}
\end{equation}
with the local Fermi momentum $k_F(r)= (3\pi^2\rho(r)/2)^{1/3}$, $M$
the nucleon mass, and $E(\vec{p}\,)= \sqrt{M^2 + \vec{p}^{\,2}}$. We
will work on an non-symmetric nuclear matter with different Fermi sea
levels for protons, $k_F^p$, than for neutrons, $k_F^n$ (equation
above, but replacing $\rho /2$ by $\rho_p $ or $\rho_n$, with
$\rho=\rho_p + \rho_n$). Finally, $A^{\mu\nu}$ is the CC nucleon
tensor~\cite{NAV04}. The $d^3p$ integrations above can be done
analytically and all of them are determined by the imaginary part of
isospin asymmetric Lindhard function, $\overline
{U}(q,k_F^n,k_F^p)$.    Explicit
expressions can be found in ~\cite{NAV04}.

\begin{figure}[htb]
 \centerline{\includegraphics[height=11cm]{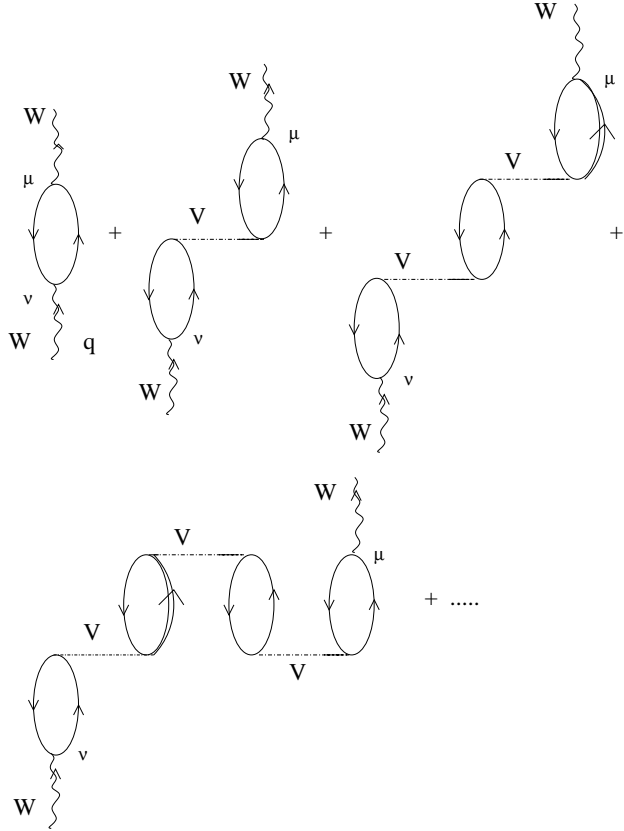}}
\vspace{-0.8cm}
 \caption{\footnotesize Set of irreducible diagrams responsible for the
  polarization (RPA) effects in the 1p1h contribution to the
  $W-$selfenergy. }\label{fig:fig3}
\end{figure}

We take into account polarization effects by substituting the particle-hole
(1p1h) response by an RPA response consisting in a series of ph and 
$\Delta$-hole excitations as shown in Fig.~\ref{fig:fig3}.
We use an effective Landau-Migdal ph-ph interaction~\cite{Sp77}:
$V = c_{0}\left\{
f_{0}+f_{0}^{\prime}\vec{\tau}_{1}\vec{\tau}_{2}+ 
g_{0}\vec{\sigma}_{1}\vec{\sigma}_{2}+g_{0}^{\prime}
\vec{\sigma}_{1}\vec{\sigma}_{2}
\vec{\tau}_{1}\vec{\tau}_{2}
\right\}$,
where only the isovector terms contribute to CC processes. 
In the $S=1=T$ channel ($\vec{\sigma} \vec{\sigma} \vec{\tau} \vec{\tau}$
operator) we use an interaction~\cite{GNO97,CO92,pion}
with explicit $\pi$ (longitudinal) and $\rho$ (transverse) exchanges.
The $\Delta(1232)$ degrees of freedom are also included in the $S=1=T$ 
channel of the RPA response. This effective interaction is non-relativistic,
and then for consistency we will neglect terms of order ${\cal O}(p^2/M^2)$
when summing up the RPA series. 
%


We ensure the correct energy balance, of both neutrino and
antineutrino CC induced process in finite nuclei, by
modifying the energy conserving $\delta$ function in
Eq.~(\ref{eq:thadron}) to account for the $Q-$value of the reaction.

We also consider the effect of the Coulomb field of the nucleus acting
on the ejected charged lepton. This is done by including
the lepton self- energy $\Sigma_C=2k^{\prime \,0}V_C(r)$ in the
intermediate lepton propagator of Fig. \ref{fig:fig1}. 


Finally, we  account for the FSI by using nucleon propagators properly
dressed with a realistic self-energy in the medium, which depends
explicitly on the energy and the momentum~\cite{FO92}.  Thus, we
rewrite the imaginary part of the Lindhard function (ph propagator) in
terms of spectral functions $S_{p,h}(\omega,\vec p;\rho)$. 

\section{RESULTS}
%

\begin{figure*}
\includegraphics[width=14cm]{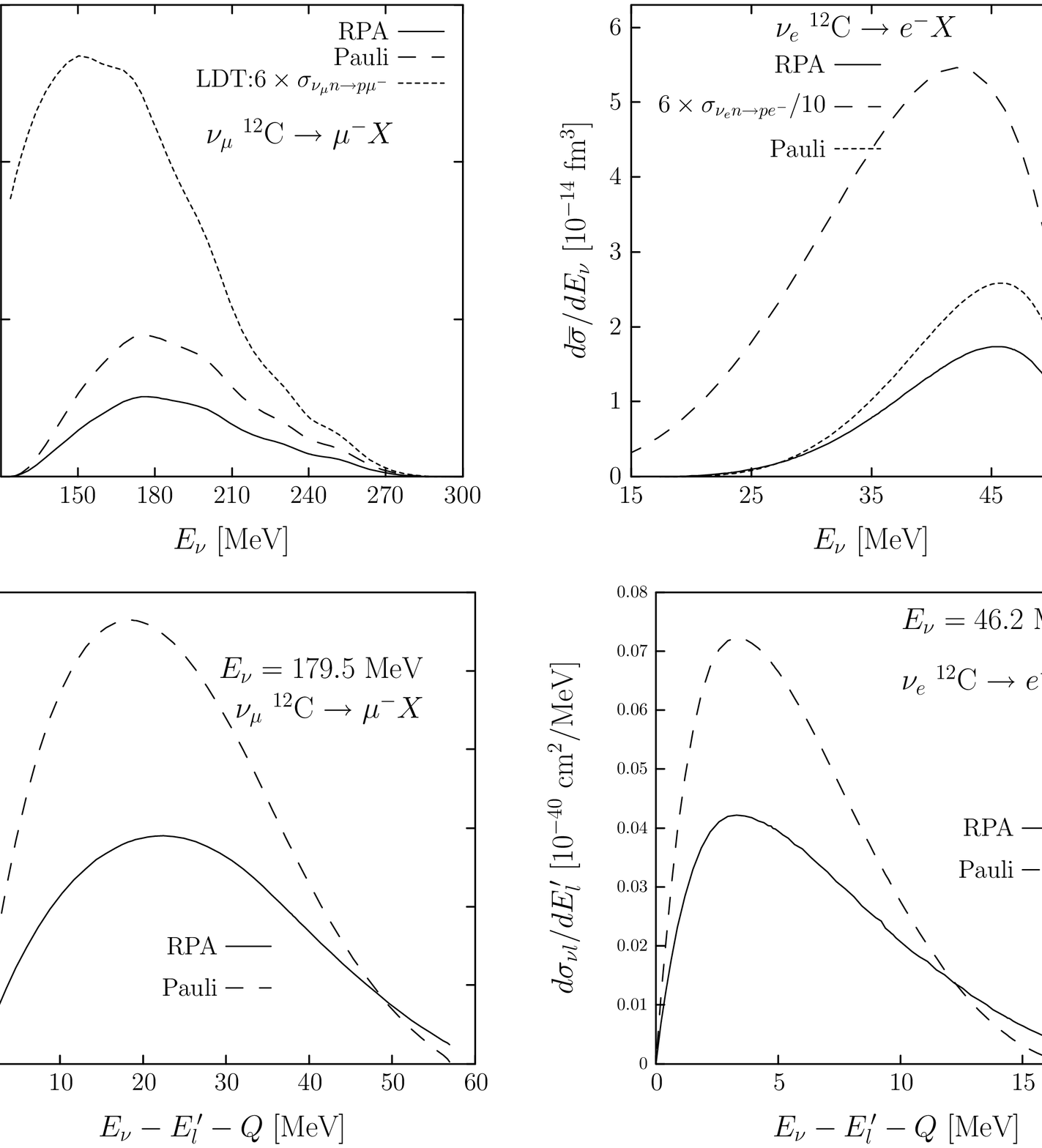}
\vspace{-3.cm}
\caption{ \footnotesize Theoretical predictions for the $^{12}$C
$(\nu_\mu,\mu^-)X$ and the $^{12}$C $(\nu_e,e^-)X$ reactions near
threshold. In addition to the full calculation (denoted as RPA), with
all nuclear effects with the exception of the FSI ones, we show
results obtained without including RPA, FSI and Coulomb corrections
(denoted as Pauli), and also results (denoted as LDT) obtained by
multiplying the free space cross section by the number of neutrons of
$^{12}$C.} \label{fig:lsnd}
\end{figure*}

\begin{table}[htb]
\caption{ \footnotesize Experimental and theoretical total muon
  capture widths for different nuclei, in $[10^4\, s^{-1}]$
  units. Data are taken from Ref.~\protect~\cite{Su87},using a
  weighted average: $\overline{\Gamma}/\sigma^2 = \sum_i
  \Gamma_i/\sigma_i^2$, with $1/\sigma^2 = \sum_i 1/\sigma_i^2$. We
  quote two different theoretical results: with (RPA) and without
  (Pauli) the inclusion of RPA correlations. FSI effects are
  not included in any of them.}
\label{tab:capres} 
\begin{tabular}{lccc}\hline\tstrut 
  & Pauli  & RPA  &
Exp   \\\hline \tstrut $^{12}$C & 5.42 & 3.21 &
$3.78\pm 0.03$  \\ $^{16}$O & 17.56 & 10.41 &
$10.24\pm 0.06$  \\ $^{18}$O & 11.94 & 7.77 & $8.80\pm 0.15$
 \\ $^{23}$Na &58.38 & 35.03 & $37.73\pm 0.14$ 
 \\ $^{40}$Ca &465.5 &257.9 &$252.5\pm 0.6 $ 
 \\ $^{44}$Ca &318 &189 & $179 \pm 4 $  \\ $^{75}$As
&1148 & 679 & 609$\pm 4$  \\ $^{112}$Cd &1825 & 1078 &
1061$\pm 9 $  \\ $^{208}$Pb & 1939 & 1310 & 1311$\pm 8 $  \\ \hline 
\end{tabular}
\end{table}
\begin{table*}
\caption{ \footnotesize Electron neutrino (left) and
  antineutrino (right)  inclusive QE  integrated cross sections from
  oxygen. We present results for relativistic ('REL') and
  non-relativistic nucleon kinematics. In this latter case, we present
  results with ('FSI') and without ('NOREL') FSI effects. Results, denoted
  as 'RPA' and 'Pauli'  have been obtained  with and
  without including RPA correlations and Coulomb corrections, respectively.
  Units: $10^{-40}$cm$^2$. }
\begin{tabular}{clcccccc}\hline\tstrut 
$E_\nu$ [MeV]&&
 \multicolumn{3}{c}{$\sigma\left(^{16}{\rm O}(\nu_e,e^-X)\right)$}&
 \multicolumn{3}{c}{$\sigma\left(^{16}{\rm O}({\bar\nu}_e,e^+X)\right)$} \\\hline\tstrut
 &  &~~ REL & ~~NOREL & FSI  & ~~REL & ~~NOREL &  FSI \\\hline\tstrut  
 400&Pauli ~&~~389.4&~~416.6&352.5&~~130.0&~~139.1&121.0\\\tstrut              
& RPA~ & ~~294.7    & ~~322.6 &303.6 &~~91.9  &~~101.9  &104.8\\\hline\tstrut  
 310   & Pauli  ~  & ~~281.4&~~297.4&240.6 &~~98.1 &~~ 104.0 & 87.2 \\\tstrut  
              & RPA                      ~   & ~~192.2    & ~~209.0 &195.2 &~~65.9 & ~~72.4  & 73.0   \\\hline\tstrut  
220           & Pauli  ~   & ~~149.5    & ~~156.2 &121.2 &~~60.7 & ~~63.6  &51.0\\\tstrut  
              & RPA                      ~   & ~~ 90.1   & ~~ 97.3  &92.8  &~~36.8 & ~~40.0  &40.2    \\\hline\tstrut  
\end{tabular}
\label{tab:fsi}
\end{table*} 
We present in Fig.~\ref{fig:lsnd} and Tables~\ref{tab:lsnd}
and~\ref{tab:eincl} our theoretical predictions and a comparison of
those to the experimental measurements of the inclusive $^{12}$C
$(\nu_\mu,\mu^-)X$ and $^{12}$C $(\nu_e,e^-)X$ reactions near
threshold.  Pauli blocking and the use of the correct
energy balance improve the results, but only once RPA and Coulomb
effects are included a good description of data is achieved.

Given the success in describing the LSND measurement of the reaction
$^{12}$C $(\nu_\mu,\mu^-)X$ near threshold, it seems natural to
further test our model by studying the closely related process of the
inclusive muon capture in $^{12}$C. Furthermore, and since there is
abundant and accurate measurements on nuclear inclusive muon capture
rates through the whole Periodic Table, we have also calculated muon
capture widths for a few selected nuclei. Results are compiled in
Table~\ref{tab:capres}. Data are quite accurate, with precisions
smaller than 1\%, quite far from the theoretical uncertainties of any
existing model.  Medium polarization effects (RPA correlations), once
more, are essential to describe the data.  Despite of the huge range
of variation of the capture widths (note, $\Gamma^{\rm exp}$ varies
from about 4$\times 10^4$ s$^{-1}$ in $^{12}$C to 1300 $\times 10^4$
s$^{-1}$ in $^{208}$Pb), the agreement to data is quite good for all
studied nuclei, with discrepancies of about 15\% at most. It is
precisely for $^{12}$C, where we find the greatest discrepancy with
experiment. Nevertheless, our model provides one of the best existing
combined description of the inclusive muon capture in $^{12}$C and the
LSND measurement of the reaction $^{12}$C $(\nu_\mu,\mu^-)X$ near
threshold.

\begin{figure}
 \vspace{-0.7cm}
 \includegraphics[width=7.7cm]{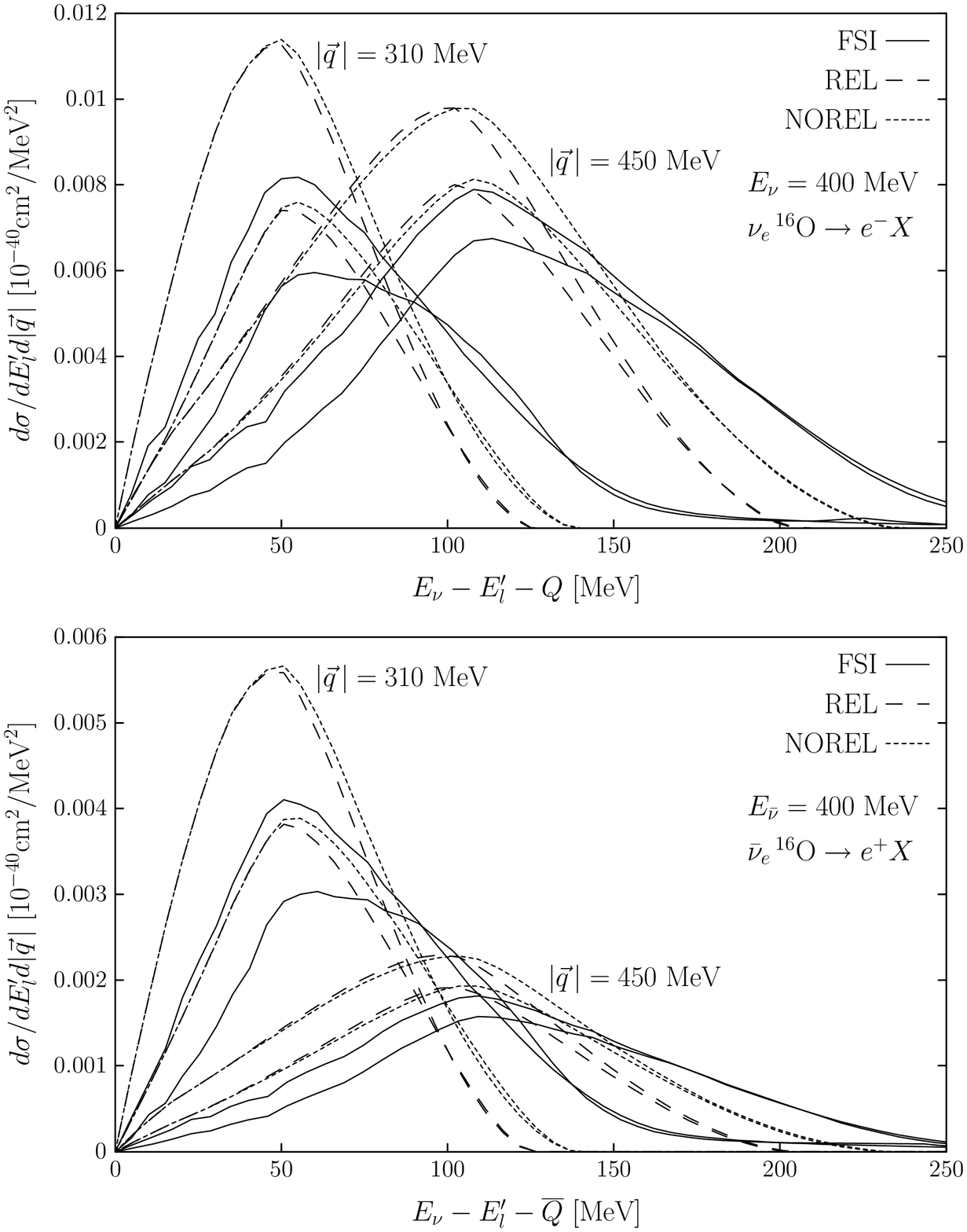}
\vspace{-1.5cm}
\caption{\footnotesize $\nu_e-$ (top) and ${\bar \nu}_e-$
  (bottom) inclusive QE differential cross sections in oxygen as a
  function of the transferred energy, at two values (310 and 450 MeV)
  of the transferred momentum.  The incoming neutrino (antineutrino)
  energy is 400 MeV. We show results for relativistic (REL) and
  non-relativistic nucleon kinematics. In this latter case, we present
  results with (FSI) and without (NOREL) FSI effects. For the three
  cases, we also show the effect of taking into account RPA
  correlations and Coulomb corrections (lower lines at the peak).}
  \label{fig:fsi}
\end{figure}

\begin{figure}
\vspace{-0.7cm}
\includegraphics[width=7.5cm]{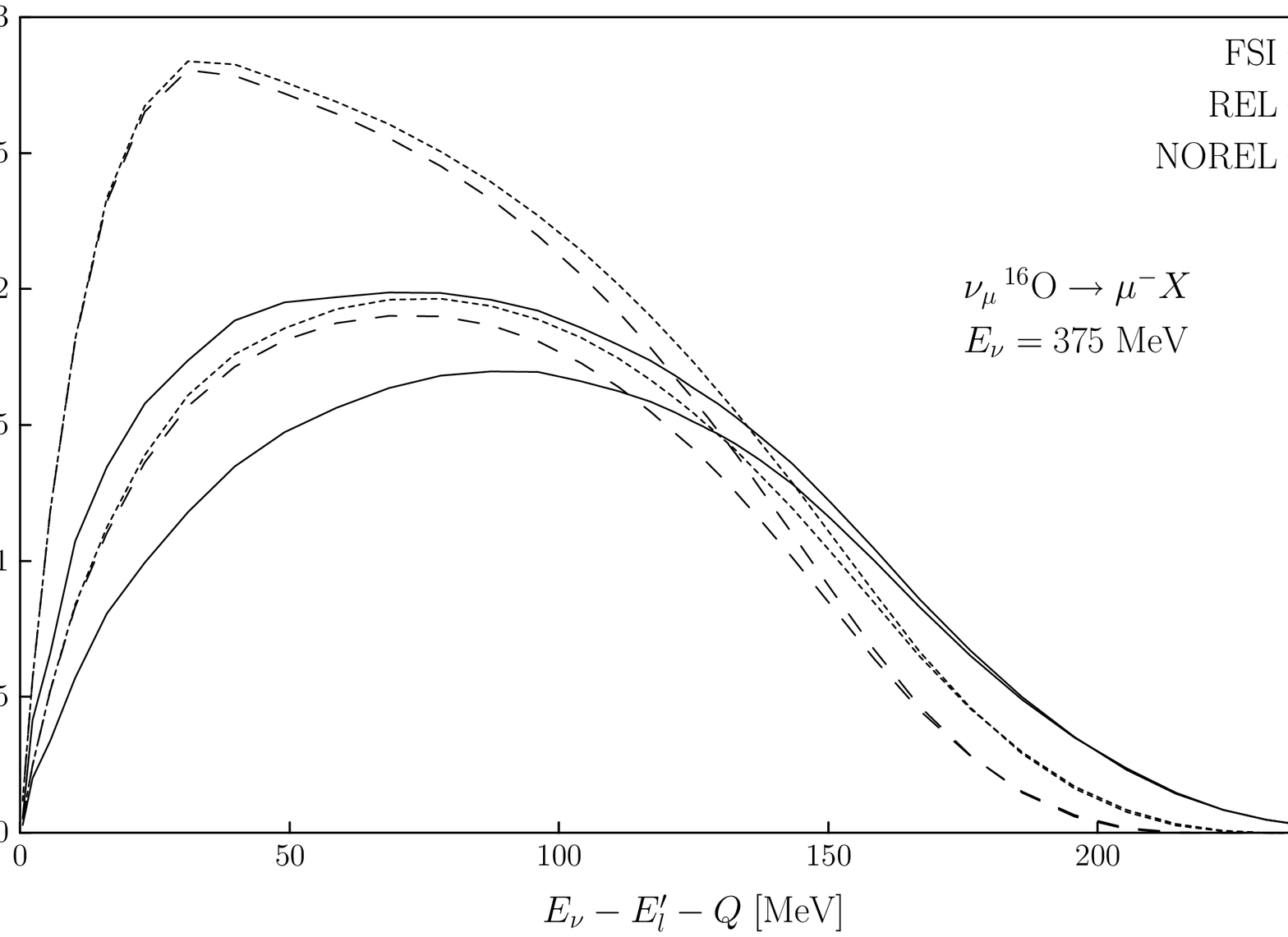}
\vspace{-5.5cm}
\caption{ \footnotesize Muon neutrino inclusive QE differential cross
  sections in oxygen as a function of the transferred energy.  The
  incoming neutrino energy is 375 MeV. The notation for the
  theoretical predictions is the same as in
  Fig.~\protect\ref{fig:fsi}.  }
  \label{fig:fsi3}
\end{figure}

 
At intermediate energies the predictions of our model become reliable
not only for integrated cross sections, but also for differential
cross sections. We present results for incoming neutrino energies
within the interval 150-400 (250-500) MeV for electron (muon)
species. In Figs.~\ref{fig:fsi} and~\ref{fig:fsi3}, FSI effects on
differential cross section are shown.  As expected, FSI provides a
broadening and a significant reduction of the strength of the QE
peak. Nevertheless the integrated cross section is only slightly
modified.  In Table~\ref{tab:fsi} we compile electron neutrino and
antineutrino inclusive QE integrated cross sections from oxygen.
Though FSI change importantly the differential cross sections, it
plays a minor role when one considers total cross sections. When
medium polarization effects are not considered, FSI provides
significant reductions (13-29\%) of the cross sections. However, when
RPA corrections are included the reductions becomes more moderate,
always smaller than 7\%, and even there exist some cases where FSI
enhances the cross sections. This can be easily understood by looking
at Fig.~\ref{fig:fsi3}. There, we see that FSI increases the cross
section for high energy transfer. But for nuclear excitation energies
higher than those around the QE peak, the RPA corrections are
certainly less important than in the peak region. Hence, the RPA
suppression of the FSI distribution is significantly smaller than the
RPA reduction of the distribution determined by the ordinary Lindhard
function.

\end{document}